# Growth of Topological Insulator Bi$_2$Se$_3$ Particles on GaAs via Droplet Epitaxy

Sivakumar Vishnuvardhan Mambakkam[1], Saadia Nasir[2], Wilder Acuna[1], Joshua M. O. Zide[1], and Stephanie Law[1, 2, a)]

[1] Department of Materials Science and Engineering, University of Delaware, 201 DuPont Hall, 127 The Green, Newark, Delaware 19716

[2] Department of Physics and Astronomy, University of Delaware, 217 Sharp Lab, 104 The Green, Newark, Delaware 19716

[a] Electronic mail: slaw@udel.edu

## Abstract

The discovery of topological insulators (TIs) and their unique electronic properties has motivated research into a variety of applications, including quantum computing. It has been proposed that TI surface states will be energetically discretized in a quantum dot nanoparticle. These discretized states could then be used as basis states for a qubit that is more resistant to decoherence. In this work, prototypical TI Bi$_2$Se$_3$ nanoparticles are grown on GaAs (001) using the droplet epitaxy technique, and we demonstrate the control of nanoparticle height, area, and density by changing the duration of bismuth deposition and substrate temperature. Within the growth window studied, nanoparticles ranged from 5-15 nm tall with an 8-18nm equivalent circular radius, and the density could be relatively well controlled by changing the substrate temperature and bismuth deposition time.

## Introduction

Topological insulators (TIs) are a class of materials which exhibit remarkable electronic properties [1, 2, 3, 4]. This is due to the presence of heavy metals, resulting in strong spin-orbit coupling. This spin-orbit coupling results in the bands overlapping, causing a rearrangement of the bands, or band inversion. This means the symmetry within both the conduction and valence bands is flipped near the Gamma point in the Brillouin zone. When TIs contact materials with a different symmetry, TI surface states form to satisfy the interfacial boundary conditions. These are electronic states which cross the bulk band gap of the TI and are physically located at the TI surfaces. Electrons occupying these states have several unique properties: they are delocalized on the surface of the TI, they are nearly massless, and they are spin-momentum locked. Altogether, this makes electrons occupying these states resistant to scattering into other surface states in the absence of a magnetic perturbation. This behavior has motivated interest in TIs in a variety of contexts such as spintronic or optoelectronic devices [5, 6].

One application for TIs is creating quantum dots to serve as room temperature quantum bits or "qubits". When reduced to nanoscale dimensions, the TI surface states are predicted to become quantized [7, 8]. These quantized states could serve as the qubit basis states. The reduced scattering pathways could then reduce qubit decoherence, potentially leading to room-temperature operation. Before creating devices, however, we must first answer two questions: 1) How do we produce the uniform TI nanoparticles needed to reduce inhomogeneous broadening and 2) Do the TI nanoparticles show evidence of quantized surface states? The work discussed in this paper focuses on answering the first question.

There are a variety of methods by which TI nanoparticles can be made. In this work, we use a growth technique called "droplet epitaxy" to produce $Bi_2Se_3$ nanoparticles [9, 10, 11]. We start by exposing the substrate to a small amount of bismuth which does not wet the substrate while keeping the substrate at a relatively low temperature. This promotes the formation of 3D particles or "droplets". Next, the particles are exposed to an overpressure selenium. These atoms then incorporate into the nanoparticles, forming $Bi_2Se_3$ since this is the most stable compound in the Bi-Se phase diagram. Droplet epitaxy can therefore be used to form pure $Bi_2Se_3$ nanoparticles with no secondary phases. Related growth studies have been performed using both metalorganic vapor phase epitaxy (MOVPE) and molecular beam epitaxy (MBE) to grow both bismuth droplets and bismuth selenide particles [12, 13, 14, 11].

In this paper, we discuss the growth of $Bi_2Se_3$ nanoparticles on GaAs via droplet epitaxy using molecular beam epitaxy (MBE). We varied the substrate temperature during bismuth deposition and the duration of bismuth exposure to understand how these parameters impact the final nanoparticle height, area, and density. The goal was to determine the extent to which TI nanoparticles could be repeatably and controllably grown. We show that the substrate temperature growth window was relatively small, approximately 35° C. Within this window, nanoparticle heights and areas were relatively similar within error bars, approximately 5-15 nm tall and 200-1000 $nm^2$ (equivalent circular radius of approximately 8-18 nm) on average. The most consistently controllable variable was nanoparticle density, which increased with increasing bismuth exposure time and decreased somewhat with increasing substrate temperature. Overall, we demonstrate control over nanoparticle dimensions but within a somewhat narrow range.

**Experimental Procedure**

Samples were synthesized using a Veeco GenXplor MBE system in the University of Delaware Materials Growth Facility. Samples were grown on epi-ready (001) GaAs wafers. Wafers were loaded into the chamber and heated to 760-770 ⁰C to desorb the oxide layer. To prevent gallium droplet formation during oxide desorption, a selenium overpressure of 7.5-9.6 x $10^{-6}$ Torr was used, as measured by beam flux monitoring (BFM). The selenium flux was started once the substrate temperature exceeded 300 ⁰C. After bringing the substrate up to 770 ⁰C, the substrate was cooled to the growth temperature. The selenium valve stayed open until the substrate dropped below 300 ⁰C. All temperatures were measured by non-contact thermocouple.

A two-step process was used to grow the selenized bismuth nanoparticles. After the substrate stabilized at the desired initial growth temperature, the bismuth shutter was opened for between 20 to 100 seconds. The bismuth cell temperature was fixed at 480 ºC, with a BFM-measured flux of 7.1-7.3 x $10^{-8}$ Torr. We exposed the substrate to a relatively small flux of bismuth, approximately 1.045 x $10^{13}$ atoms/($cm^2$s), to form 3-dimensional bismuth nanoparticles. Next, the bismuth shutter was closed, and the sample was annealed for between 100 to 20 seconds, such that the total time of bismuth deposition plus annealing is fixed at 120 seconds. Finally, the substrate temperature was lowered to 50 ºC (ramp rate of 20 ºC/min.) and the selenium shutter was opened, with a BFM flux of 3.5-5.3 x $10^{-6}$ Torr, forming $Bi_2Se_3$ nanoparticles. Selenium is kept open as the substrate cools and is closed once the substrate cools below 200 ºC. The duration of selenium exposure varies, with growths done at 250 ºC taking 6-7 minutes to cool, and growths done at 275/285 ºC taking 9-10 minutes to cool. After selenium is closed, the sample is removed from the chamber for analysis.

Two parameters were varied to tune the nanoparticle area, height, and density: the duration of bismuth deposition ($t_{Bi}$) and the temperature of the substrate used for initial bismuth deposition ($T_{sub}$). After growth, samples were examined using atomic force microscopy (AFM) and scanning electron microscopy (SEM). AFM was conducted using the Dimension-3100 V SPM system in the Keck Center for Advanced Microscopy and Microanalysis, and SEM was done using the Zeiss Merlin SEM system in the UD Nanofabrication Facility cleanroom. Quantities of interest for our study are nanoparticle height (by AFM), area (by SEM), and nanoparticle density per unit area (by AFM).

**Results and Discussion**

In this work, we studied growths done within the $T_{sub}$ window of 250 to 285 ºC, and with $t_{Bi}$ between 20 and 100 seconds. As discussed in detail below, we discovered that substrates mounted on different plates experienced different actual temperatures even when the thermocouple read the same temperature. For this reason, we indicate the sample plate (Q2 or Q3) used for each sample. We will present data for a variety of pairs of samples grown with different bismuth deposition times or with different temperatures, then synthesize all results to form a picture of the dynamics of $Bi_2Se_3$ nanoparticle formation.

We first explored the properties of nanoparticles as a function of bismuth deposition time. Figure 1 shows data for the growths done at $T_{sub}$=285 ºC: A (C) and B (D) are the AFM (SEM) images for growths done with $t_{Bi}$=20s and 100s, respectively. E and F summarize the statistical height and area data extracted from the AFM and SEM images, respectively. This information is presented in the form of box plots, which summarize multiple statistical quantities about the entire dataset. All box plots presented here include the following quantities: the box bounds the 25th to 75th percentile data, the line cutting through the box is the median, the solid square shows the mean, and the whiskers extend to the farthest datapoint that falls within the range of the Lower Inner and Upper Inner Fences (defined by $25^{th}$ percentile minus 1.5*interquartile range, and $75^{th}$ percentile plus 1.5*interquartile range, respectively). Solid diamonds above or below the plot represent outliers, some of which may be cut out of the plot for ease of viewing. See supplementary material at [URL will be inserted by AIP Publishing] for details on how AFM and SEM images were processed to obtain numerical data, as well as histograms of the datasets. AFM images were used to measure nanoparticle heights and density, and SEM images were used to measure nanoparticle area. Area data from AFM images was not used due to error caused by tip-sample convolution which may cause particles to seem larger in area than they are. The densities of nanoparticles per unit area, as measured by AFM, is $1.53 \times 10^{-5}$ particles/nm$^2$ for $t_{Bi}=20$s and $5.13 \times 10^{-5}$ particles/nm$^2$ for $t_{Bi}=100$s (3.35 times more). For this pair of samples, within error bars, we see little to no change in average nanoparticle height. However, the sample with 100 seconds of bismuth deposition shows a notable increase in polydispersity, 1.6x increase in average area, and a 3x increase in nanoparticle density. By polydispersity, we refer to the relative number of nanoparticles which deviate from the average for a specific property, as well as the degree of their deviation. This can be expressed qualitatively by the change in the size of the box in the box plots or expressed quantitatively by statistical quantities like the standard deviation.

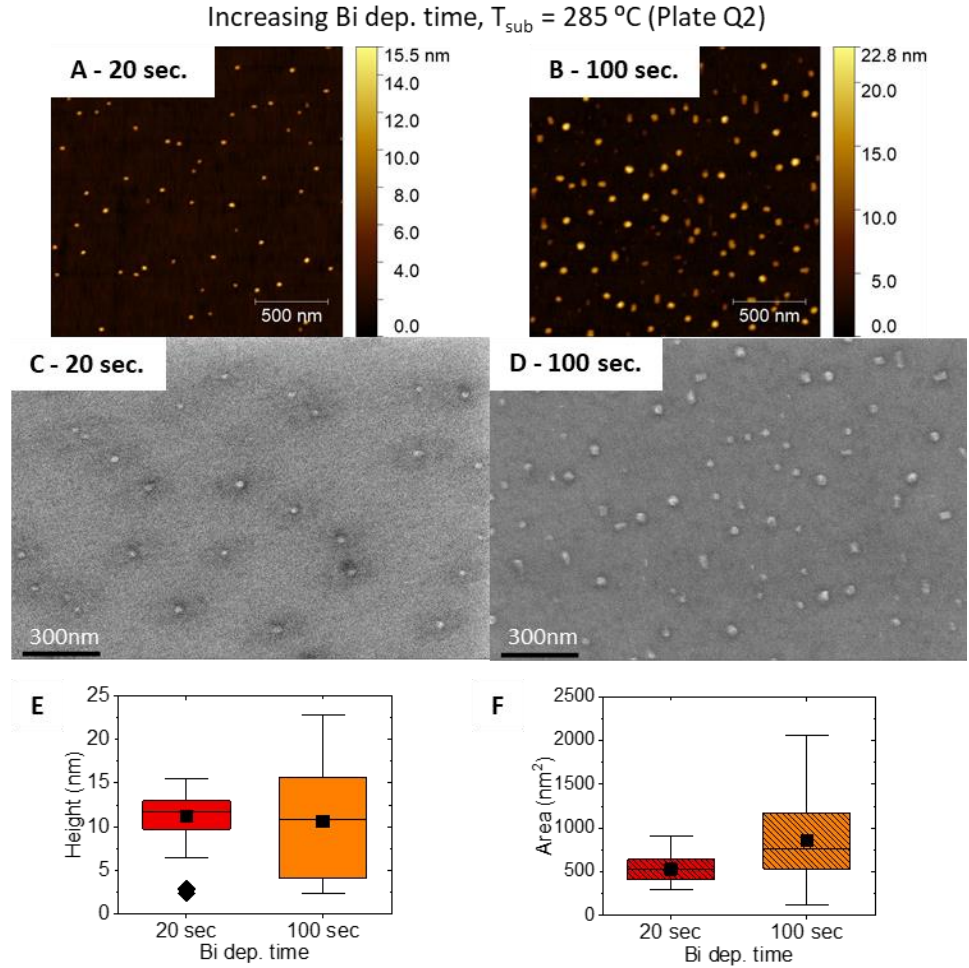

*Figure 1: AFM and SEM images and resultant box plots for growths done at a $T_{sub}$ of 285 ºC with varying $t_{Bi}$. A and C are the AFM and SEM images for the growth done with 20 sec. $t_{Bi}$ (grown 10/03/2020), B and D are the AFM and SEM images for the growth done with 100 sec. of $t_{Bi}$ (grown 10/01/2020). E and F are the box plots showing height data from the AFM and area data from the SEM images respectively. Color scales and y-axes of box plots adjusted for ease of comparison (outliers in box plot may be cut out as a result).*

Next, we compare four samples grown at $T_{sub}$=250 ºC. Figure 2 shows the following: A (E), B (F), C (G) and D (H) are the AFM (SEM) images for samples with $t_{Bi}$=20s ,60s, 80s, and 100s, respectively. Fig. 2I and J summarize the statistical height and area data extracted from the AFM and SEM images, respectively. Density values for the growths are as follows: $t_{Bi}$=20s: 2.18 x $10^{-5}$ $nm^{-2}$; $t_{Bi}$=60s: 12.8 x $10^{-5}$ $nm^{-2}$; $t_{Bi}$=80s: 18.1 x $10^{-5}$ $nm^{-2}$; and $t_{Bi}$=100s: 21.5 x $10^{-5}$ $nm^{-2}$. We note that the growth with $t_{Bi}$=20s had the selenium flux stopped at 255 ºC as opposed to 300 ºC.

Second, the AFM in Figure 2C has an adjusted scale-bar for ease of comparison (original range 0

to 36.3 nm due to an outlier). Third, the contrast in the SEM for Figure 2E was insufficient to perform an area analysis, and so this is omitted from the box plot. For this series, we continue to see a monotonic increase in density with bismuth deposition time, while nanoparticle height and area become progressively smaller.

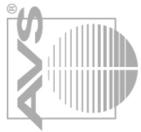

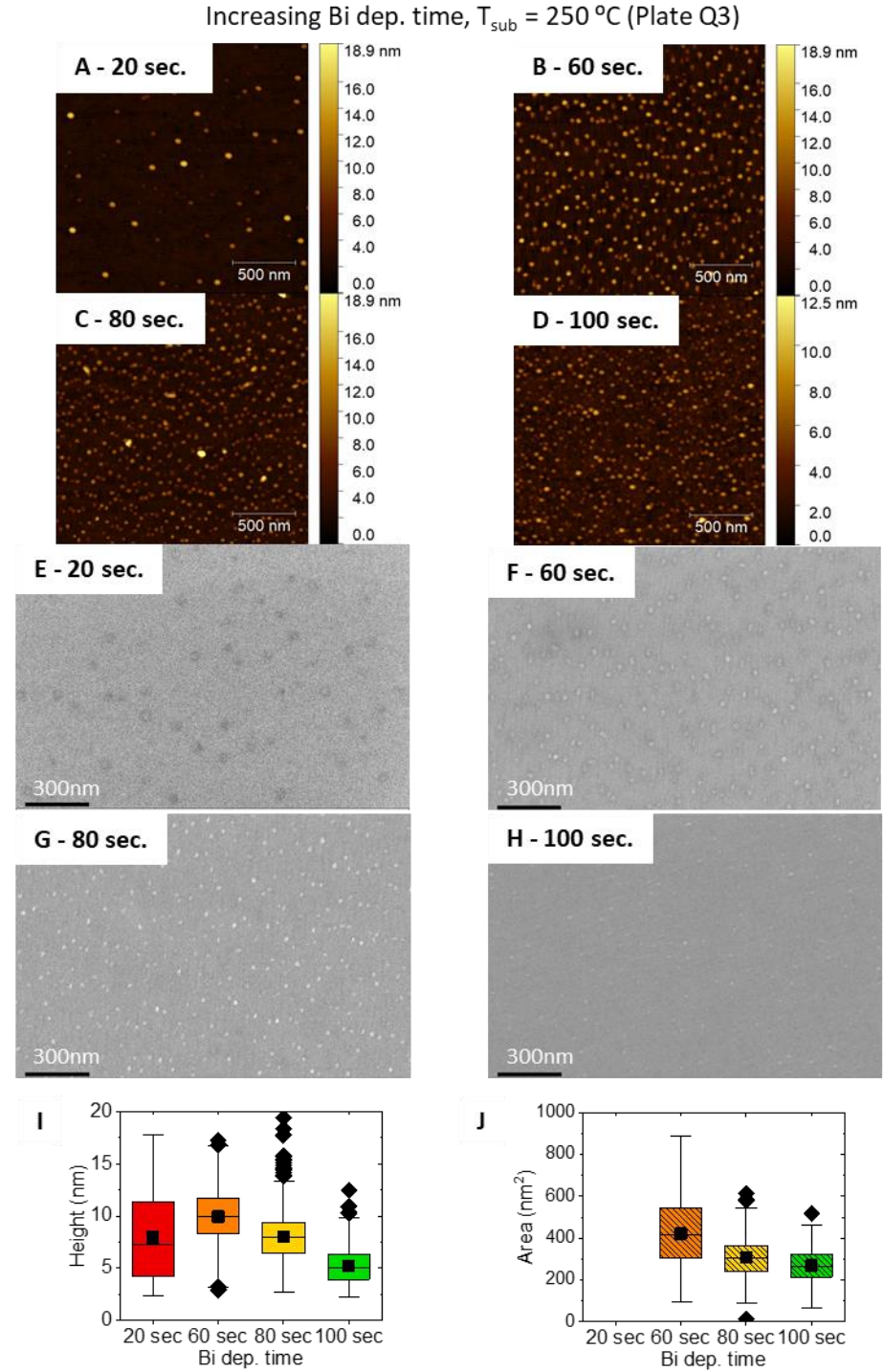

*Figure 2: AFM and SEM images and resultant box plots for growths done at a $T_{sub}$ of 250 ºC with varying $t_{Bi}$. A)/E) are the AFM and SEM images for the growth done with 20 sec. $t_{Bi}$ (grown 02/26/2021), B)/F) are the AFM and SEM images for the growth done with 60 sec. of $t_{Bi}$ (grown 03/08/2021), C)/G) are the AFM and SEM images for the growth done with 80 sec. of $t_{Bi}$ (grown 02/14/2021), D)/H) are the AFM and SEM images for the growth done with 100 sec. of $t_{Bi}$ (grown 03/23/2021). I) and J) are the box plots showing height data from the AFM and area data from the SEM images respectively. Color scales and y-axes of box plots adjusted for ease of comparison (outliers in box plot may be cut out as a result).*

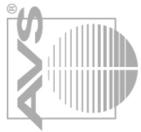

We now examine the effect of changing substrate temperature with a constant bismuth deposition time. We start with two growths done with $t_{Bi}$=100s, but with a substrate temperature difference of 10ºC. Figure 3 shows data for the growths done with $t_{Bi}$=100s: A (C) and B (D) are the AFM (SEM) images corresponding to growths done at 275 and 285 ºC, respectively. The nanoparticle density for the growth at $t_{Bi}$=100s and $T_{sub}$=275 ºC is 4.33 x $10^{-5}$ $nm^{-2}$, and for the growth done at $T_{sub}$=285 ºC is 5.13 x $10^{-5}$ $nm^{-2}$ (1.18 times more). The nanoparticle area, height, polydispersity, and density for both samples are the same to within error bars.

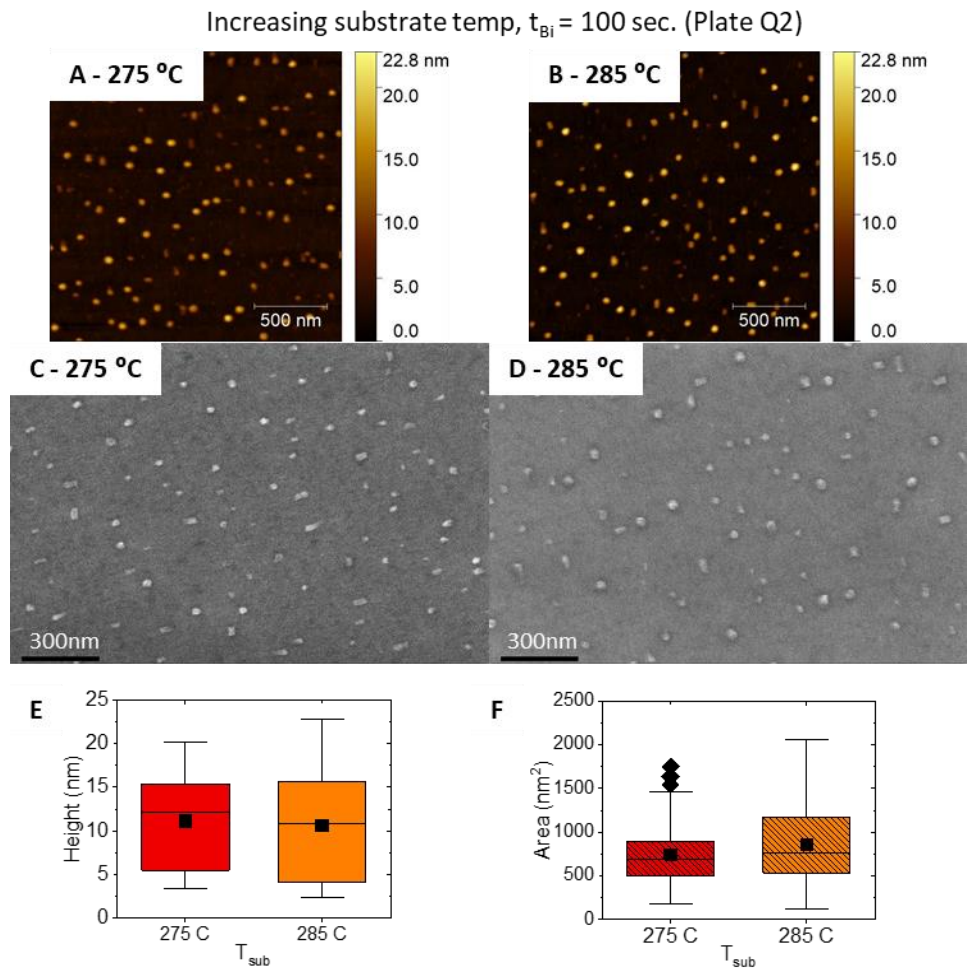

*Figure 3: AFM and SEM images and resultant box plots for growths done with 100 sec. $t_{Bi}$ and varying substrate temperature. A)/C) are the AFM and SEM images for the growth done at a $T_{sub}$ of 275 ºC (grown 09/25/2020), B)/D) are the AFM and SEM images for the growth done at a $T_{sub}$ of 285 ºC (grown 10/01/2020). E) and F) are the box*

*plots showing height data from the AFM and area data from the SEM images respectively. Color scales and y-axes of box plots adjusted for ease of comparison (outliers in box plot may be cut out as a result).*

Now we look to another pair of growths, this time with $t_{Bi}$=60s, but with a larger temperature difference of 25 ºC. Figure 4 shows the following: A (C) and B (D) are the AFM (SEM) images corresponding to growths done at 250 and 275 ºC, respectively. The nanoparticle density for the growth at $T_{sub}$=250 ºC is 8.98 x $10^{-5}$ $nm^{-2}$, and density for the growth done at $T_{sub}$=275 ºC is 5.33 x $10^{-5}$ $nm^{-2}$ (0.59 times less). Unlike the previous case, we see a more significant change with the substrate temperature change. Average particle height increases by 1.43x and average particle area increases by 1.98x with increasing temperature.

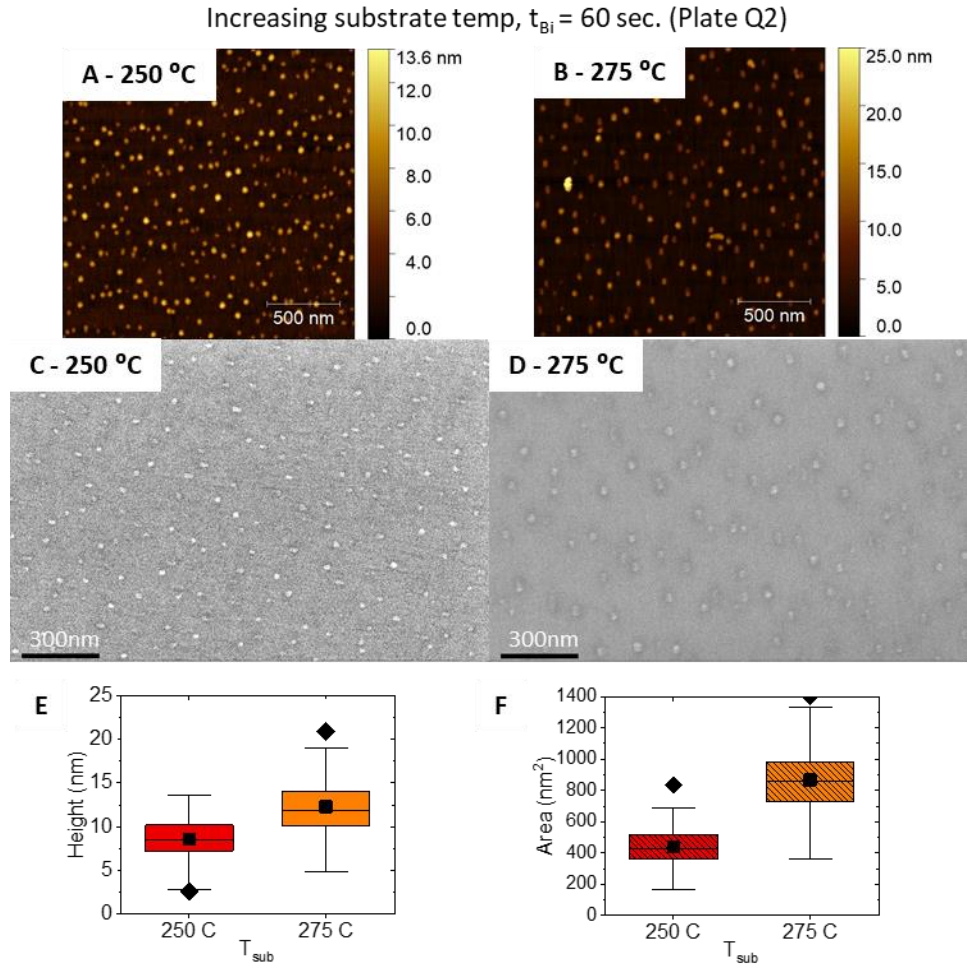

*Figure 4: AFM and SEM images and resultant box plots for growths done with 60 sec. $t_{Bi}$ and varying substrate temperature (plate Q2). A)/C) are the AFM and SEM images for the growth done at a $T_{sub}$ of 250 ºC (grown*

Finally, we examine a series of three growths with $t_{Bi}$=60s, grown at 250ºC, 275ºC, and 285ºC done on plate Q3. Figure 5 shows the following: A (D), B (E), and C (F) correspond to the AFM (SEM) images for the growths done at 250 ºC, 275 ºC, and 285 ºC, respectively. Note that the scalebar in Figure 5B was adjusted for ease of comparison. Density for the growth at 250 ºC is 12.8 x 10⁻⁵ nm⁻², 275 ºC is 11.9 x 10⁻⁵ nm⁻², and 285 ºC is 4.45 x 10⁻⁵ nm⁻². We again observe an increase in average nanoparticle area and height with increasing substrate temperature.

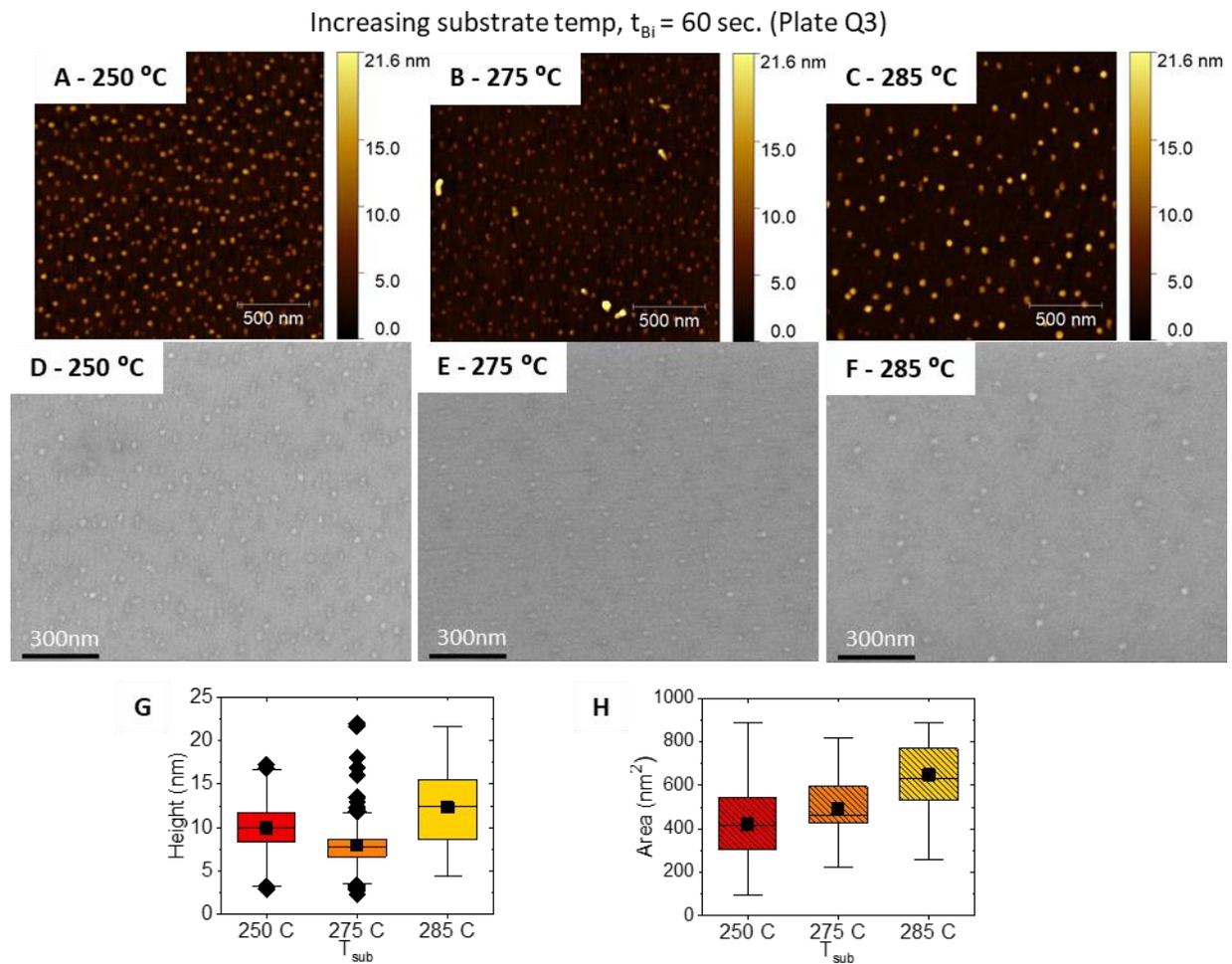

*Figure 5: AFM and SEM images and resultant box plots for growths done with 60 sec. $t_{Bi}$ and varying substrate temperature (plate Q3). A)/D) are the AFM and SEM images for the growth done at a $T_{sub}$ of 250 ºC (grown 03/08/2021), B)/E) are the AFM and SEM images for the growth done at a $T_{sub}$ of 275 ºC (grown 03/10/2021), and*

*C)/F) are the AFM and SEM images for the growth done at a $T_{sub}$ of 285 ºC (grown 03/11/2021). G) and H) are the box plots showing height data from the AFM and area data from the SEM images respectively. SEM images were cropped. See supplementary material at [URL will be inserted by AIP Publishing] for full scans. Color scales and y-axes of box plots adjusted for ease of comparison (outliers in box plot may be cut out as a result).*

Attempts were made to synthesize bismuth nanoparticles at both higher and lower substrate temperatures. The AFM scans for two attempts made at 225 ºC and one at 325 ºC are shown in Figure 6. All three samples showed a rough surface, and except for a single feature on Figure 6C and some short (<5 nm tall) features in Figure 6B, we did not show evidence of nanoparticle formation.

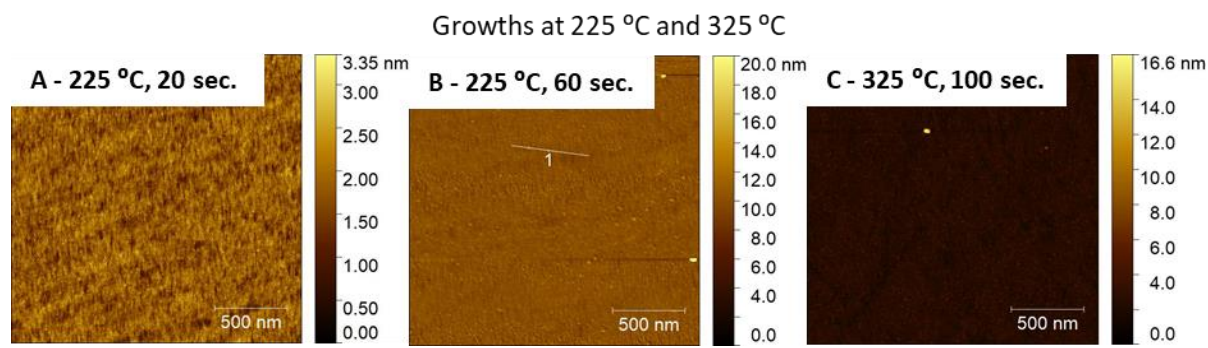

*Figure 6: AFM images of growths performed at $T_{sub}$'s of 225 and 325 ºC. A) is from a growth performed with 20 sec. $t_{Bi}$ at a $T_{sub}$ of 225 ºC (plate Q3, 08/07/2020), B) is from a growth performed with 60 sec. $t_{Bi}$ at a $T_{sub}$ of 225 ºC (plate Q2, 08/07/2020), and C) is from a growth performed with 100 sec. $t_{Bi}$ at a $T_{sub}$ of 325 ºC (plate Q3, 12/17/2020). Color scales and y-axes of box plots adjusted for ease of comparison (outliers in box plot may be cut out as a result).*

## Discussion

We begin by discussing the series in Figures 1 and 2, which show trends of particle height and area with varying $t_{Bi}$. Neither of the series at 285 ºC or 250 ºC show changes in particle height and area within error bars as $t_{Bi}$ changes. However, both show significant increases in particle density as $t_{Bi}$ increases. This suggests that bismuth atoms upon impinging onto the surface are more likely to nucleate new particles as opposed to incorporate into existing ones. This trend is more dramatic for the growths done at lower substrate temperature, which supports this theory, as the mobility of the atoms on the substrate decreases with decreasing temperature, leading to more nucleation events.

Next, we consider the series in Figures 3, 4, and 5, which show trends of particle height and area with varying $T_{sub}$. In Figure 3, the series with a 10 ºC difference, we see little change in particle properties, whereas in Figure 4, the series with a 25 ºC difference, an increase in particle height and area is observed along with a decrease in particle density at higher temperatures. This is also seen in the full series from 250 to 285 ºC shown in Figure 5. We note here the main difference between the growths in Figure 3 versus those in Figures 4 and 5 is the amount of bismuth supplied: the growths in Figure 3 had $t_{Bi}$=100s, whereas the growths in Figures 4 and 5 all had $t_{Bi}$=60s. Overall, we observe that as the temperature increases, the nanoparticle height and area usually increase and the density decreases. This is not surprising. Since the amount of bismuth deposited is the same, if the particle area and height increase, the density must decrease. For higher temperatures, bismuth adatoms have a longer diffusion length, thus enabling the formation of larger particles rather than the nucleation of new particles. This density reduction with increasing substrate temperature was also observed by C. Li for bismuth droplets on GaAs grown via MBE [14]. However, this change in particle properties with increasing substrate temperature may also depend on the amount of bismuth deposited.

Now we consider the results for the growths done at 225 ºC and 325 ºC shown in Figure 6. In these growths, the only evidence of nanoparticle formation was a few short features in Figure 6B and one feature in Figure 6C. We therefore concluded that the substrate temperature window to achieve nanoparticle formation is small, less than 100ºC. It is possible that the particles become increasingly more sparse on the substrate as the temperature increases and so it becomes increasingly difficult to detect them. This is potentially supported by recent AFM results made on the sample shown in Figure 6C, where taller, larger-area nanoparticle-like features (~100-120 nm tall, ~100 nm equivalent circular radii) are observed when increasing the scan size to 5x5 μm. Further work to identify the nature of these features is necessary, due to their large deviation from the features typically observed. Regardless, a more detailed study of the growth space between 285ºC to 325 ºC is needed.

Next, we will discuss the issue of reproducibility. Figure 7 shows two examples of repeated growths. The growths shown in Figure 7A and B were performed with $t_{Bi}$=100s at $T_{sub}$=250ºC (on 09/25/2020 and 03/23/2021), and Figure 7D and E were performed with $t_{Bi}$=60s at $T_{sub}$=285ºC (on 10/03/2020 and 03/11/2021). Figure 7C and F are the respective height data box plots for each pair of growths showing the change with time. We observe a difference in nanoparticle height and area for the samples shown in Fig. 7A and B, indicating poor reproducibility, while we see similar nanoparticle height and area distributions for the samples shown in Fig. 7C and D, indicating good reproducibility.

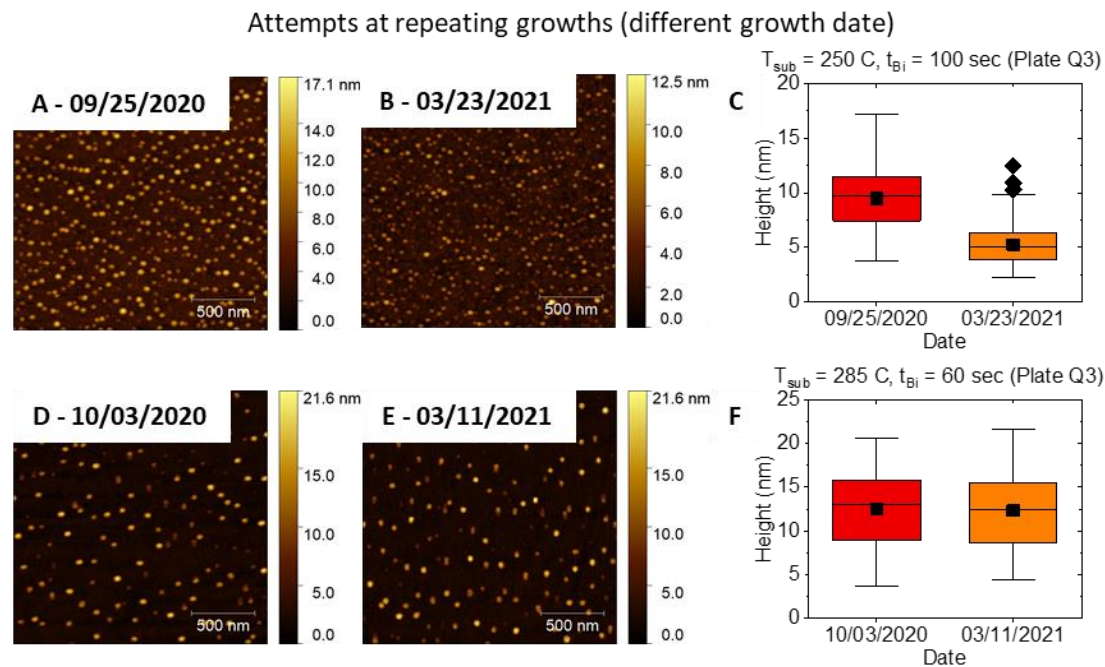

*Figure 7: AFM images of repeated growths (grown on different days). A)/B) are for growths performed with 100 sec. $t_{Bi}$ at a $T_{sub}$ of 250 °C (plate Q3), D)/E) are for growths performed with 60 sec $t_{Bi}$ at a $T_{sub}$ of 285 °C (plate Q3). C)/F) are box plots showing height data for the pairs of A)/B) and D)/E) respectively. Color scales and y-axes of box plots adjusted for ease of comparison (outliers in box plot may be cut out as a result).*

We attribute difficulties in reproducibility to the substrate holder becoming coated in selenium over time, causing a different physical substrate temperature for a similar thermocouple reading. This interpretation is supported by the difference in nanoparticle properties for growths performed with the same parameters, but on different substrate holders. We show two examples of this in Figure 8, with growths done on nearby growth days, but on different sample plates. The samples shown in Figure 8A and B were grown on 02/26/2021 and 03/08/2021 respectively, with the sample shown in Figure 8A grown on plate Q2 and the sample shown in Figure 8B grown on plate Q3. Figure 8C is the height data box plot for the pair of growths in Figures 8A and B. Nanoparticle densities for these growths are 8.98 x 10^{-5} nm^{-2} and 12.8 x 10^{-5} nm^{-2} for the samples shown in Figure 8A and B respectively. The samples shown in Figure 8D and E were grown on 03/11/2021 and 03/10/2021 respectively, with the sample shown in Figure 8D grown on plate Q2 and the sample shown in Figure 8E grown on plate Q3. Figure 8F is the height data box plot for

the pair of growths in Figures 8D and E. Nanoparticle densities for these growths are 5.33 x $10^{-5}$

$nm^{-2}$ and 11.9 x $10^{-5}$ $nm^{-2}$ for the samples shown in Figures 8D and E respectively. From both the

box plots and the density values, there is a clear difference in properties for growths done using different substrate holders. This has also been discussed in other growth studies published by our group [15].

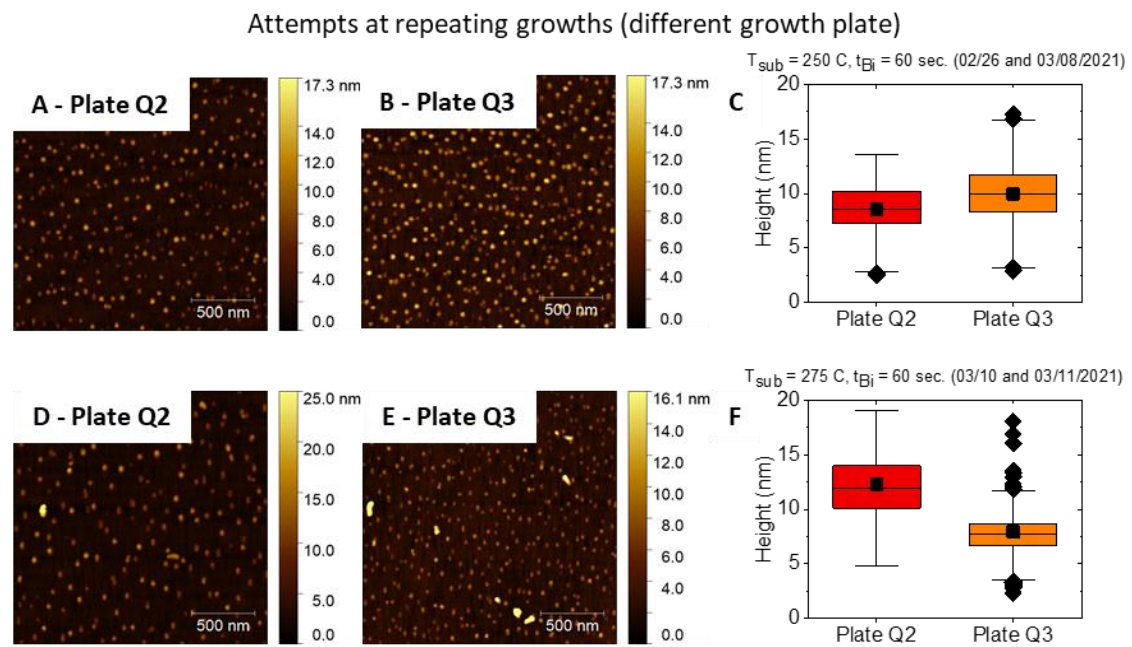

Figure 8: AFM images of repeated growths. A)/B) are for growths performed with 60 sec $t_{Bi}$ at a $T_{sub}$ of 250 ºC (plate Q2 on 02/26/2021, plate Q3 on 03/08/2021), D)/E) are for growths performed with 60 sec. $t_{Bi}$ at a $T_{sub}$ of 275 ºC (plate Q2 on 03/11/2021, plate Q3 on 03/10/2021). C)/F) are box plots showing height data for the pairs of A)/B) and D)/E) respectively. Color scales and y-axes of box plots adjusted for ease of comparison (outliers in box plot may be cut out as a result).

To estimate the deviation in physical temperature between sample plates Q2 and Q3, we used the temperature at which the oxide desorbed from our GaAs wafers which should always be the same physical temperature. We measured this temperature by heating the sample in increments of 5ºC while using reflection high-energy electron diffraction (RHEED) to monitor the surface reconstruction. GaAs deoxidation occurs when the RHEED pattern transitions from rings to streaks. Using this method, we approximate the deviation between plate Q2 and Q3 to be between 0 and 10º C. This agrees with the data shown in Fig. 7. As shown in the previous section, growths done at 225 ºC showed little to no evidence of nanoparticles. These growths were done on 08/07/2020. We have also presented data for samples grown at 250 ºC which were gathered in the range 02/26/2021 to 03/23/2021, and these growths all showed clear presence of nanoparticles. We therefore believe that the shift in physical temperature due to successive growths is < 25 ºC. We note that we have few datapoints for this deoxidation temperature and a relatively high uncertainty. In addition, the temperature deviation at the high temperatures of the deoxidation point (710-730 ºC by thermocouple) may not be representative of the deviation at lower temperatures. However, we are confident that a change in physical temperature caused the difficulties in reproducibility. This highlights the extreme sensitivity of these samples to substrate temperature as well as the relatively narrow substrate temperature window.

Finally, we contrast the growth procedure and results presented in this work with previously published results. In particular, the growths of bismuth nanoparticles by C. Li et al. [14] and $Bi_2Se_3$ nanoparticles by M. Claro et al. [11] will be discussed. Both works use the technique of droplet epitaxy to form bismuth nanoparticles on GaAs, with the latter work extending the recipe to include selenium exposure to form $Bi_2Se_3$ nanoparticles. Both works provided a proof of concept for creating nanoparticles of the kind we were interested in studying, and our approach was adapted from the ones discussed in these works.

The first point of contrast in our approach is the preparation of the initial GaAs substrate. As purchased, GaAs wafers have an epi-ready oxide layer on top which is desorbed at a high temperature prior to growth. This desorption is typically done under an overpressure of arsenic, to reduce the formation of gallium droplets and the surface roughness. Afterward, a GaAs overlayer is often grown to further improve the quality of the growth surface. This is the approach used in both aforementioned works.

However, for MBE growth chambers where selenium is used as a source, it is not recommended to have an arsenic source in the same chamber as there is the risk of arsenic substitution and incorporation in selenide materials due to their similar atomic radii. Therefore, for each growth described here the substrate would have to be transferred into an arsenide growth chamber for oxide desorption and GaAs growth, after which it would be transferred into the selenide growth chamber for the nanoparticle growth. This is the approach used in [11]. For the volume of growths required in our study, this approach was deemed to not be feasible in the short term. Therefore, as described in the Experimental Procedure section, we desorbed the epi-ready oxide under an overpressure of selenium, and this is the growth surface which was used for the growth of our nanoparticles.

This approach is not very well-studied. One study of the passivation of GaAs surfaces with selenium suggests that selenium can incorporate to form a stable surface [16], although their method for preparing the surface is not the same as what is done during desorption of the epi-ready oxide. Preliminary XPS measurements of our surfaces suggest that there is some selenium incorporation. We do also observe droplet-like features in SEM on a larger size scale than the nanoparticle studied in this work. See supplementary material at [URL will be inserted by AIP Publishing] for more details on these as well as other features observed in the growths aside from the nanoparticles studied.

We expect that this difference in substrate preparation method plays a role in nanoparticle formation. We note that while the previously mentioned studies discuss growths with (1x3) and c(4x4) reconstructions of GaAs, we observe a (2x4) reconstruction using our method. We suspect that both the surface roughness and the number of dangling bonds on the surface are different for the surfaces prepared using a selenium overpressure. This would in turn affect the diffusion and nucleation of bismuth nanoparticles by changing the bonding at the substrate/nanoparticle interface. Specifically, with a possibly greater number of dangling bonds and several unsatisfied selenium bonds on the substrate surface, bismuth atoms may tend towards nucleating new particles by bonding with the selenium. This is at least partially supported by our observation of increased nanoparticle density with bismuth deposition time. This theory of stronger bonding at the nanoparticle/substrate interface could also explain why our attempts to remove these nanoparticles from GaAs substrates via sonication of the samples in solvents have failed, despite their nature as van der Waals materials. To verify this theory, a detailed TEM study of the nanoparticle/substrate interface is needed.

We also performed one growth in which the GaAs oxide layer was desorbed under an arsenic overpressure and a GaAs layer was grown, followed by transfer of the substrate under vacuum into our selenide growth chamber to perform the $Bi_2Se_3$ nanoparticle growths. We show in Figure 9 the AFM and SEM images for this growth, which was performed using a substrate temperature of 250°C and a bismuth deposition time of 60 seconds (all other parameters such as flux and approximate Bi:Se flux ratio are as described in the Experimental Procedure section). We observe nanoparticles similar to those previously reported, with most between 10-20nm tall and with effective radii between 25-45nm. Compared to the nanoparticles grown on the selenium-prepared surfaces, these are generally taller, larger in area, and significantly less dense. The difference in morphology and density can likely be attributed to a smoother surface after the growth of the GaAs layer compared to the samples deoxidized under a selenium overpressure. A smoother surface will have fewer defects at which particles can nucleate as well as longer adatom diffusion lengths.

### Growth of nanoparticles on smoother GaAs

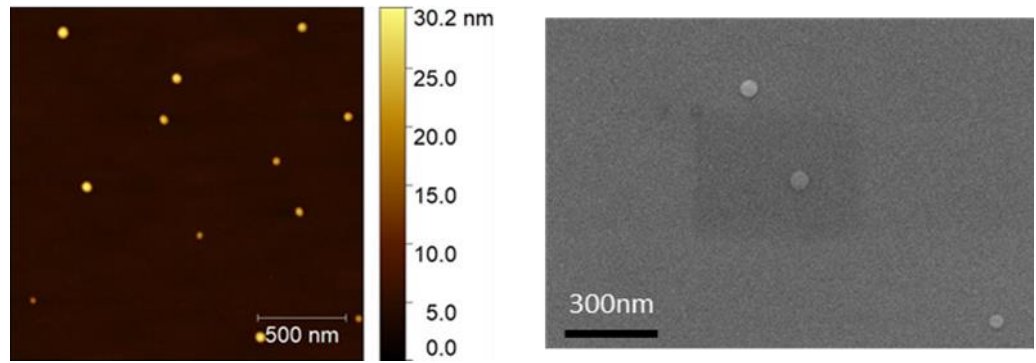

*Figure 9: AFM and SEM images of growths performed on GaAs prepared by desorption under arsenic overpressure. Growths performed with 60 seconds of bismuth deposition at 250 ºC.*

Another deviation between our growths and previous reports are the exact details of the growth procedure. In principle, the growth of $Bi/Bi_2Se_3$ nanoparticles by droplet epitaxy is quite simple, and just involves the deposition of bismuth under conditions that promote nanoparticle formation, followed by (for the formation of $Bi_2Se_3$) sufficient selenium exposure to cause incorporation and crystallization. We summarize the growth procedures for the three works discussed so far in Table 1.

*Table 1: Growth procedures for Bi/Bi2Se3 nanoparticles grown on GaAs(001) by MBE*

| Report / Growth step | C. Li et al. (Bi nanoparticles) | M. Claro et al. (Bi and $Bi_2Se_3$ nanoparticles) | Our growth recipe |
|---|---|---|---|
| **Bismuth deposition** | Stabilization at substrate temperature | Stabilization at substrate temperature, | Stabilization at substrate temperature for between |
| | for 30 minutes, then deposition with fixed bismuth flux for between 10 to 35 seconds (for fixed substrate temperature), and for 15 seconds (for varying substrate temperature). | then deposition with fixed bismuth flux until GaAs RHEED pattern was no longer visible (approximately 15 minutes). | 11 and 27 minutes, then deposition with a fixed bismuth flux. Total time for this step is kept constant at 120 seconds, with bismuth shutter open for between 20 to 100 seconds and bismuth shutter closed for the remainder of the 120 seconds. |
| **Selenium deposition** | N/A | Closed bismuth shutter, and immediately opened selenium shutter at the same substrate temperature for 40 minutes. | After 120 seconds "bismuth step", selenium shutter opened and substrate temperature setpoint set to 50$^\circ$C. Selenium shutter closed when substrate temperature reaches 200$^\circ$C. |

In this study, a different approach is used for both the bismuth and selenium deposition steps. For the bismuth deposition, a time of 120 seconds is fixed for the "bismuth" step wherein the bismuth shutter is opened for between 20 and 100 seconds and then is closed for the remainder of the 120 seconds. Therefore, there is a period of between 20 to 100 seconds following bismuth deposition where there is no flux exposed on the substrate while it sits at the initial substrate temperature. This was done based on the assumption that total growth time would play a role in determining the final characteristics of the nanoparticles, since while the substrate is being heated the motion of atoms on the surface is inevitable. Therefore, to study the variable of total bismuth deposition in isolation, we decided to perform the growth in this way to decouple it from the variable of total growth time as much as possible.

For the selenium deposition step, the selenium shutter was opened while at the same time the substrate temperature setpoint was set to 50$^{\circ}$C. The purpose of this was to set the setpoint low enough such that the PID controlling the power outputted to the substrate heater would decrease the output to 0%, stopping heating to the substrate. We assume thus that the motion of the previously deposited bismuth becomes sufficiently slowed and the main process occurring during this time is the diffusion and incorporation of selenium into the bismuth nanoparticles. The selenium shutter is closed when the substrate temperature reaches 200$^{\circ}$C. This takes between 6-9 minutes depending on the initial substrate temperature. It is common practice in our group based on experience growing $Bi_2Se_3$ thin films to maintain a selenium overpressure while the substrate is above 200$^{\circ}$C to prevent selenium desorption due to its high vapor pressure, and thus this was done for our nanoparticle growths as well.

The final aspects to compare are the growth results. Our nanoparticles are roughly the same height as both the bismuth and bismuth selenide nanoparticles mentioned in both reports (roughly 5-15nm), although the bismuth nanoparticles discussed in the work of M. Claro et al. prior to performing selenization were somewhat taller (~29nm). In terms of area/equivalent diameter, our nanoparticles are in a similar range if not somewhat smaller than those previously reported (<36nm, as opposed to ~50nm in diameter reported in previous work). However, our estimates of nanoparticle diameter are based on SEM measurements, previous reports relied on AFM results. As AFM tips themselves are roughly 10nm in diameter, the sample-tip convolution could lead to the deviation between our results and previously-reported results. As for observed trends, C. Li et al. studied the same trends of varying bismuth deposition time and substrate temperature. Like our study, they observed a decrease in nanoparticle density with increasing substrate temperature, accompanied by an increase in nanoparticle height. However, unlike our study, they did not observe a change in nanoparticle density with varying bismuth deposition time.

**Conclusion**

In this work, we showed that $Bi_2Se_3$ nanoparticles can be self-assembled on GaAs substrates using droplet epitaxy. We demonstrated control over the nanoparticle height, area, and density by controlling the bismuth deposition time and the substrate temperature. In general, the nanoparticles showed relatively small polydispersity, meaning that most particles were of the same area and height. We found that nanoparticles will only form in a relatively small substrate temperature window. These results show that while control over the nanoparticle size is possible, there are limits on how large the particles can become using this method. For the nanoparticle size range presented (5-15 nm tall and 8-18 nm effective circular radius), theory indicates an energy level separation of ~0.05eV [8]. However, if the nanoparticle dimension is less than ~7nm, the wavefunction of the electrons occupying the surface states can tunnel through the nanoparticle and interfere with other quantized states [7]. The small size of the nanoparticles obtained by this method results in the largest energy separation of the quantized states, thus improving our chances of measuring them.

While larger particles may be possible, this would require changing one of the variables not accounted for in this study, for example moving outside of this substrate temperature window, changing the incoming bismuth flux, or changing the substrate or its preparation. We note that our group has attempted the production of similarly sized nanoparticles by means of e-beam lithography [17]. To achieve similar nanoparticle areas, heights, and densities requires a great deal of lithography time, and by extension expenditure. Therefore, this study shows the potential for creating TI nanoparticles over a large area via a relatively simple and inexpensive procedure, which will prove useful for future studies in harnessing TI surface states in devices.

**Acknowledgements**

S. V. M. and S. L. acknowledge funding from the National Science Foundation, Division of Materials Research under Award No. 1838504 and the Brookhaven National Laboratory/University of Delaware Seed Program under Award No. 20A00145. S. N. acknowledges funding from the U.S. Department of Energy, Office of Science, Office of Basic Energy Sciences, under Award No. DE-SC0017801. This research was partially supported by NSF through the University of Delaware Materials Research Science and Engineering Center DMR-2011824. The authors acknowledge the use of the Materials Growth Facility (MGF) at the University of Delaware, which is partially supported by the National Science Foundation Major Research Instrumentation under Grant No. 1828141 and UD-CHARM, a National Science Foundation MRSEC under Award No. DMR-2011824.

**Data Availability**

Data supporting the work presented is available from corresponding authors upon reasonable request.

## Tables

*Table 1: Growth procedures for Bi/Bi2Se3 nanoparticles grown on GaAs(001) by MBE*

| Report / Growth step | C. Li et al. (Bi nanoparticles) | M. Claro et al. (Bi and $Bi_2Se_3$ nanoparticles) | Our growth recipe |
|---|---|---|---|
| **Bismuth deposition** | Stabilization at substrate temperature for 30 minutes, then deposition with fixed bismuth flux for between 10 to 35 seconds (for fixed substrate temperature), and for 15 seconds (for varying substrate temperature). | Stabilization at substrate temperature, then deposition with fixed bismuth flux until GaAs RHEED pattern was no longer visible (approximately 15 minutes). | Stabilization at substrate temperature for between 11 and 27 minutes, then deposition with a fixed bismuth flux. Total time for this step is kept constant at 120 seconds, with bismuth shutter open for between 20 to 100 seconds and bismuth shutter closed for the remainder of the 120 seconds. |
| **Selenium deposition** | N/A | Closed bismuth shutter, and immediately opened selenium shutter at the same substrate temperature for 40 minutes. | After 120 seconds "bismuth step", selenium shutter opened and substrate temperature setpoint set to 50°C. Selenium shutter closed when substrate temperature reaches 200°C. |

## Figure captions

Figure 1: AFM and SEM images and resultant box plots for growths done at a $T_{sub}$ of 285 ºC with varying $t_{Bi}$. A and C are the AFM and SEM images for the growth done with 20 sec. $t_{Bi}$ (grown 10/03/2020), B and D are the AFM and SEM images for the growth done with 100 sec. of $t_{Bi}$ (grown 10/01/2020). E and F are the box plots showing height data from the AFM and area data from the SEM images respectively. Color scales and y-axes of box plots adjusted for ease of comparison (outliers in box plot may be cut out as a result).

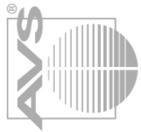

Figure 2: AFM and SEM images and resultant box plots for growths done at a Tsub of 250 ⁰C with varying tBi. A)/E) are the AFM and SEM images for the growth done with 20 sec. tBi (grown 02/26/2021), B)/F) are the AFM and SEM images for the growth done with 60 sec. of tBi (grown 03/08/2021), C)/G) are the AFM and SEM images for the growth done with 80 sec. of tBi (grown 02/14/2021), D)/H) are the AFM and SEM images for the growth done with 100 sec. of tBi (grown 03/23/2021). I) and J) are the box plots showing height data from the AFM and area data from the SEM images respectively. Color scales and y-axes of box plots adjusted for ease of comparison (outliers in box plot may be cut out as a result).

Figure 3: AFM and SEM images and resultant box plots for growths done with 100 sec. tBi and varying substrate temperature. A)/C) are the AFM and SEM images for the growth done at a Tsub of 275 ⁰C (grown 09/25/2020), B)/D) are the AFM and SEM images for the growth done at a Tsub of 285 ⁰C (grown 10/01/2020). E) and F) are the box plots showing height data from the AFM and area data from the SEM images respectively. Color scales and y-axes of box plots adjusted for ease of comparison (outliers in box plot may be cut out as a result).

Figure 4: AFM and SEM images and resultant box plots for growths done with 60 sec. tBi and varying substrate temperature (plate Q2). A)/C) are the AFM and SEM images for the growth done at a Tsub of 250 ⁰C (grown 02/26/2021), B)/D) are the AFM and SEM images for the growth done at a Tsub of 275 ⁰C (grown 03/11/2021). E) and F) are the box plots showing height data from the AFM and area data from the SEM images respectively. Color scales and y-axes of box plots adjusted for ease of comparison (outliers in box plot may be cut out as a result).

Figure 5: AFM and SEM images and resultant box plots for growths done with 60 sec. tBi and varying substrate temperature (plate Q3). A)/D) are the AFM and SEM images for the growth done at a Tsub of 250 ⁰C (grown 03/08/2021), B)/E) are the AFM and SEM images for the growth done at a Tsub of 275 ⁰C (grown 03/10/2021), and C)/F) are the AFM and SEM images for the growth done at a Tsub of 285 ⁰C (grown 03/11/2021). G) and H) are the box plots showing height data from the AFM and area data from the SEM images respectively. SEM images were cropped. See supplementary material at [URL will be inserted by AIP Publishing] for full scans. Color scales and y-axes of box plots adjusted for ease of comparison (outliers in box plot may be cut out as a result).

Figure 6: AFM images of growths performed at Tsub's of 225 and 325 ⁰C. A) is from a growth performed with 20 sec. tBi at a Tsub of 225 ⁰C (plate Q3, 08/07/2020), B) is from a growth performed with 60 sec. tBi at a Tsub of 225 ⁰C (plate Q2, 08/07/2020), and C) is from a growth performed with 100 sec. tBi at a Tsub of 325 ⁰C (plate Q3, 12/17/2020). Color scales and y-axes of box plots adjusted for ease of comparison (outliers in box plot may be cut out as a result).

Figure 7: AFM images of repeated growths (grown on different days). A)/B) are for growths performed with 100 sec. tBi at a Tsub of 250 ⁰C (plate Q3), D)/E) are for growths performed with 60 sec tBi at a Tsub of 285 ⁰C (plate Q3). C)/F) are box plots showing height data for the pairs of A)/B) and D)/E) respectively. Color scales and y-axes of box plots adjusted for ease of comparison (outliers in box plot may be cut out as a result).

Figure 8: AFM images of repeated growths. A)/B) are for growths performed with 60 sec  tBi at a Tsub of 250 ⁰C (plate Q2 on 02/26/2021, plate Q3 on 03/08/2021), D)/E) are for growths performed with 60 sec.  tBi at a Tsub of 275 ⁰C (plate Q2 on 03/11/2021, plate Q3 on 03/10/2021). C)/F) are box plots showing height data for the pairs of A)/B) and D)/E) respectively. Color scales and y-axes of box plots adjusted for ease of comparison (outliers in box plot may be cut out as a result).

Figure 9: AFM and SEM images of growths performed on GaAs prepared by desorption under arsenic overpressure. Growths performed with 60 seconds of bismuth deposition at 250 ⁰C.

Increasing Bi dep. time, $T_{sub}$ = 285 °C (Plate Q2)

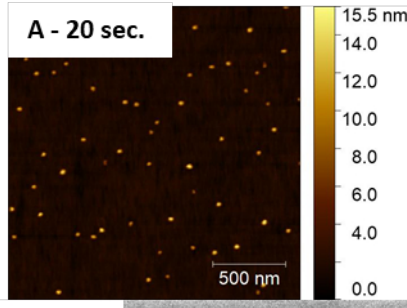

A – 20 sec.

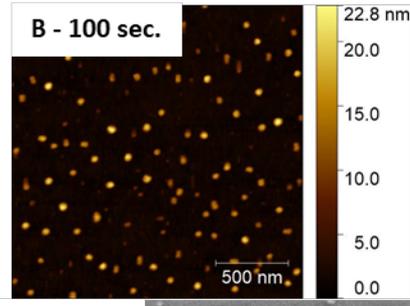

B – 100 sec.

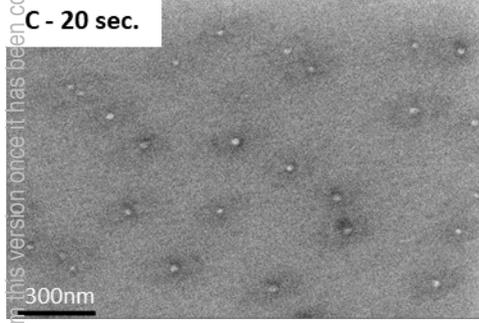

C – 20 sec.

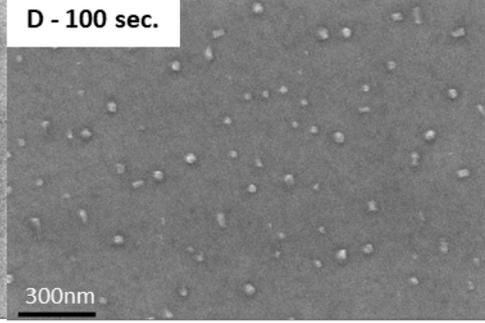

D – 100 sec.

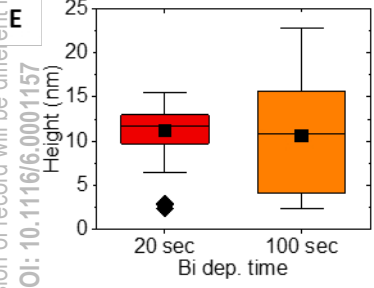

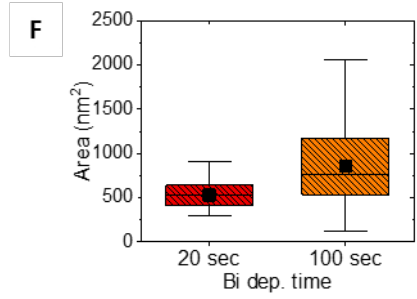

Increasing Bi dep. time, $T_{sub}$ = 250 °C (Plate Q3)

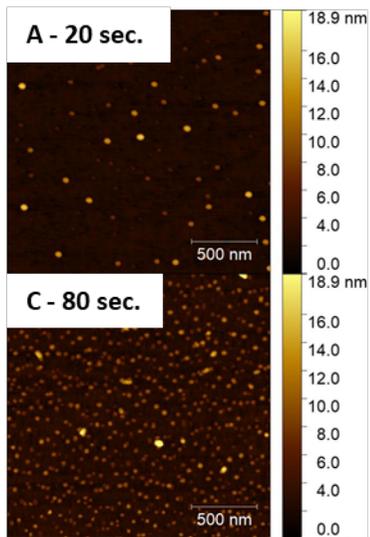

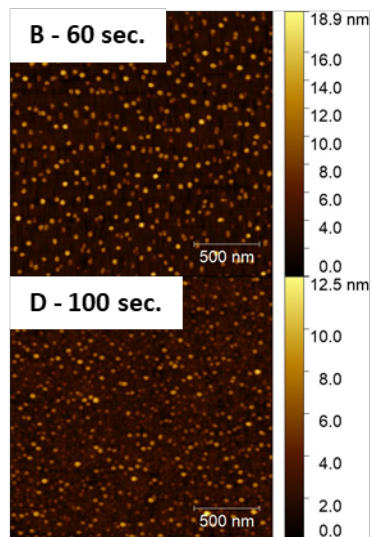

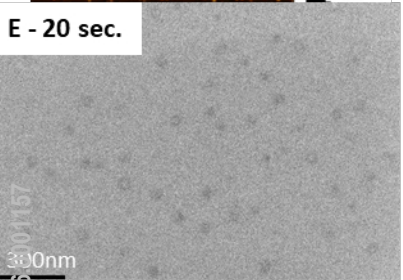

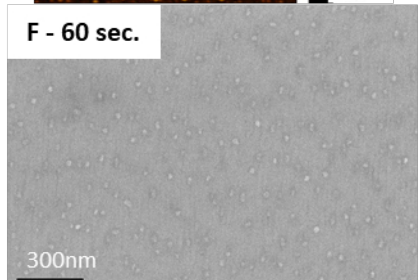

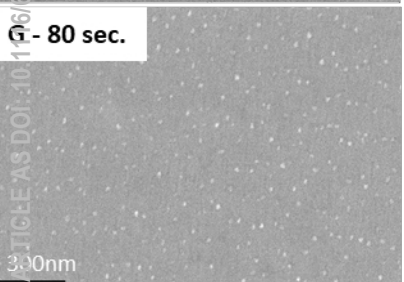

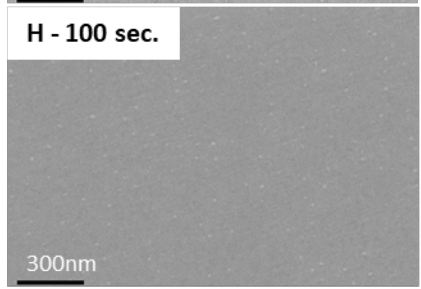

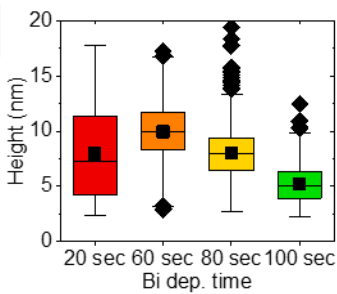

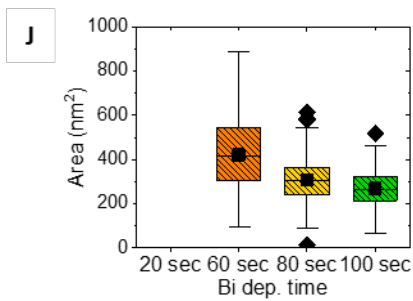

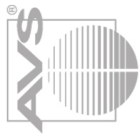

Increasing substrate temp, $t_{Bi}$ = 100 sec. (Plate Q2)

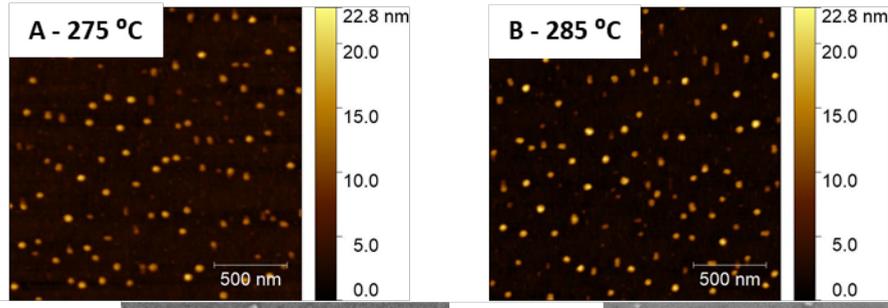

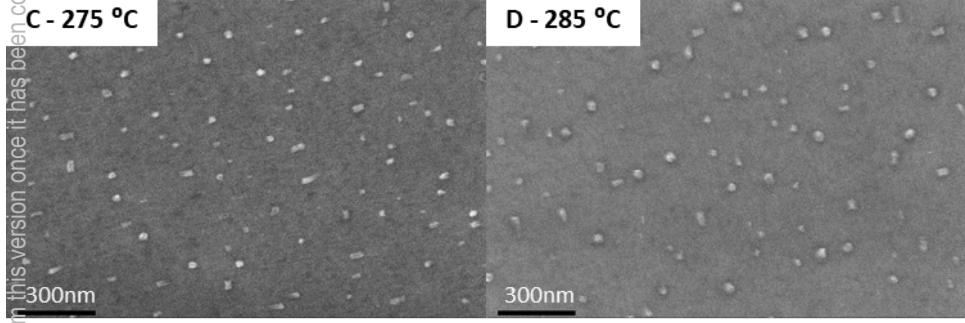

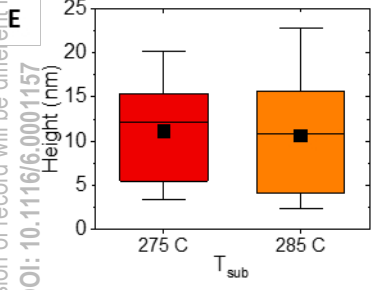

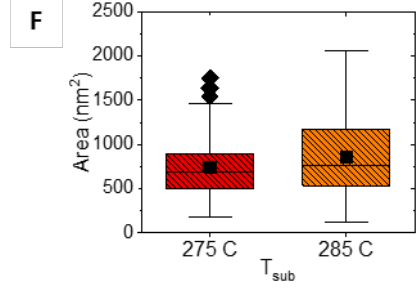

Increasing substrate temp, $t_{Bi}$ = 60 sec. (Plate Q2)

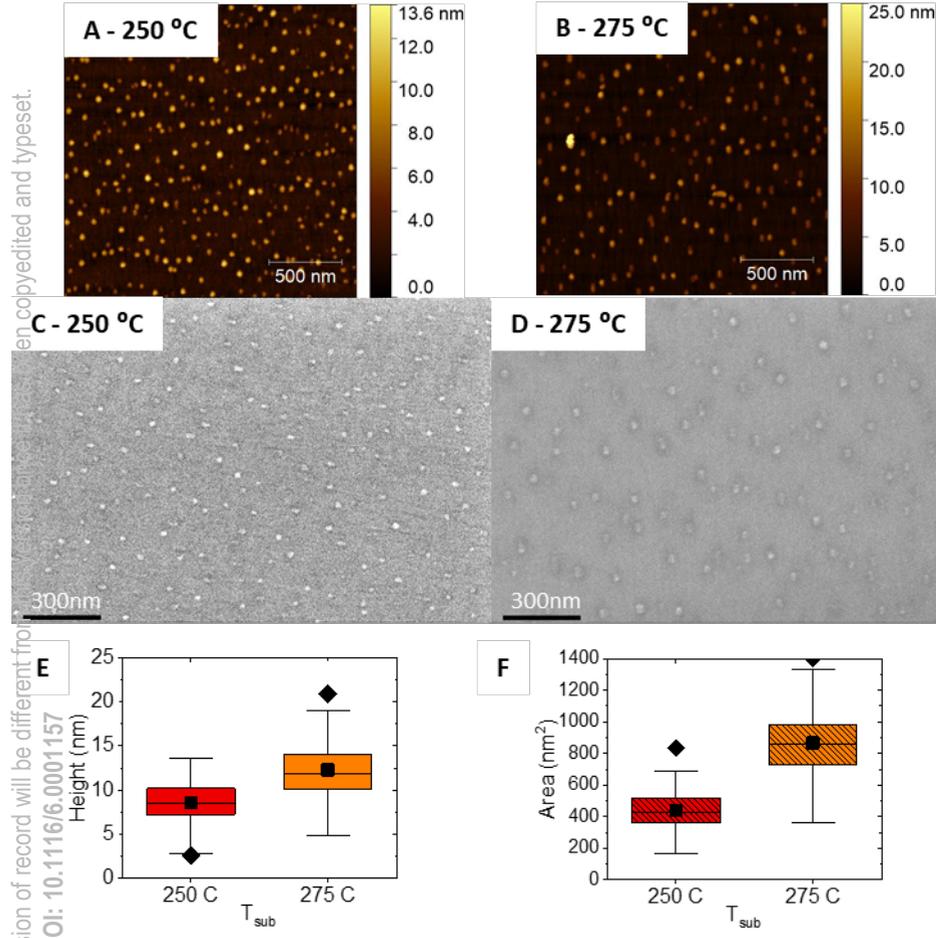

Increasing substrate temp, $t_{Bi}$ = 60 sec. (Plate Q3)

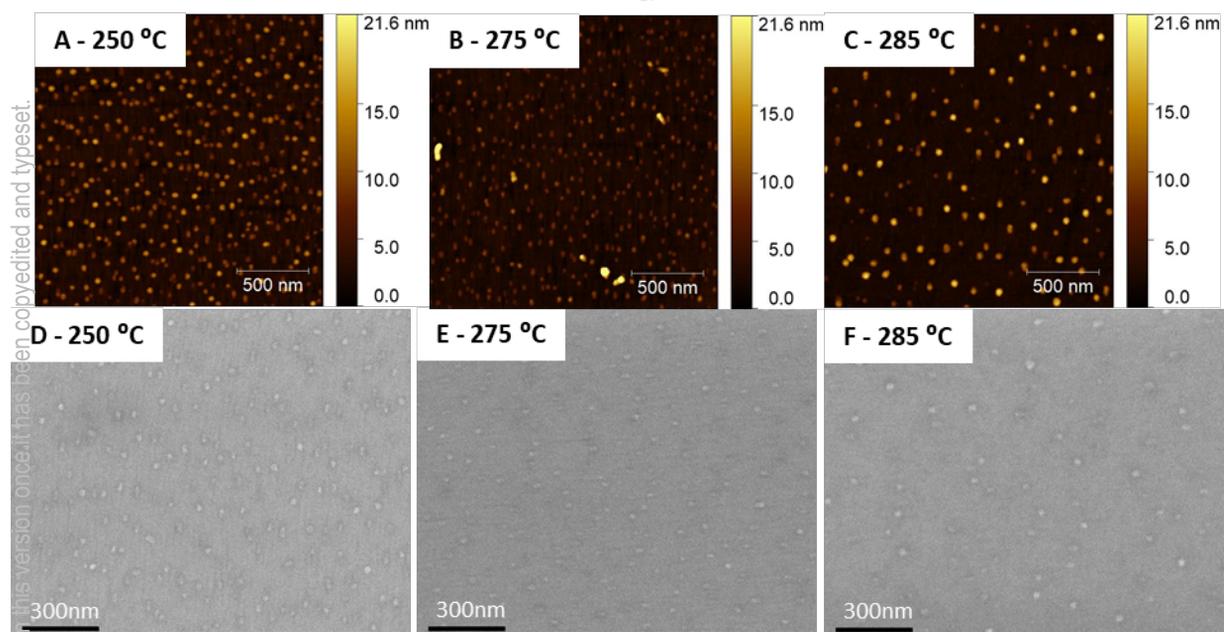

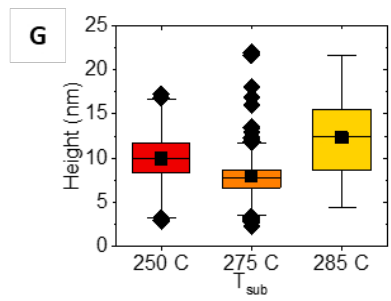

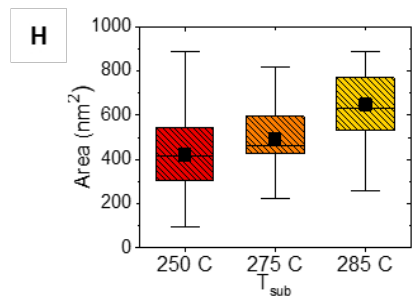

Growths at 225 ºC and 325 ºC

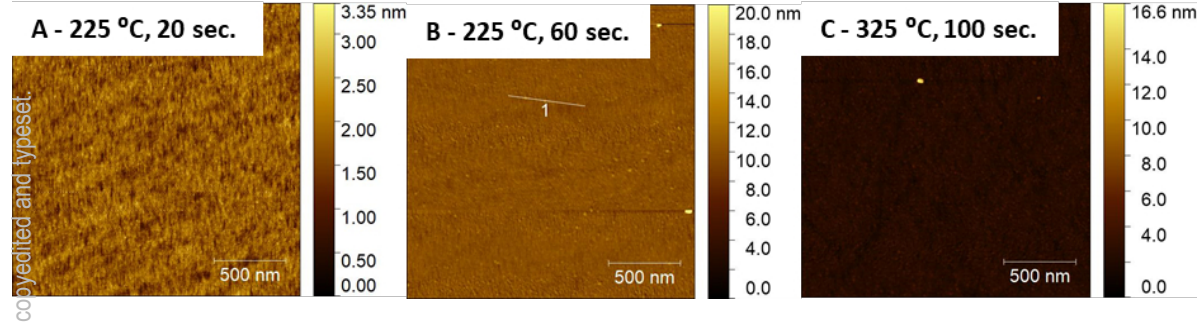

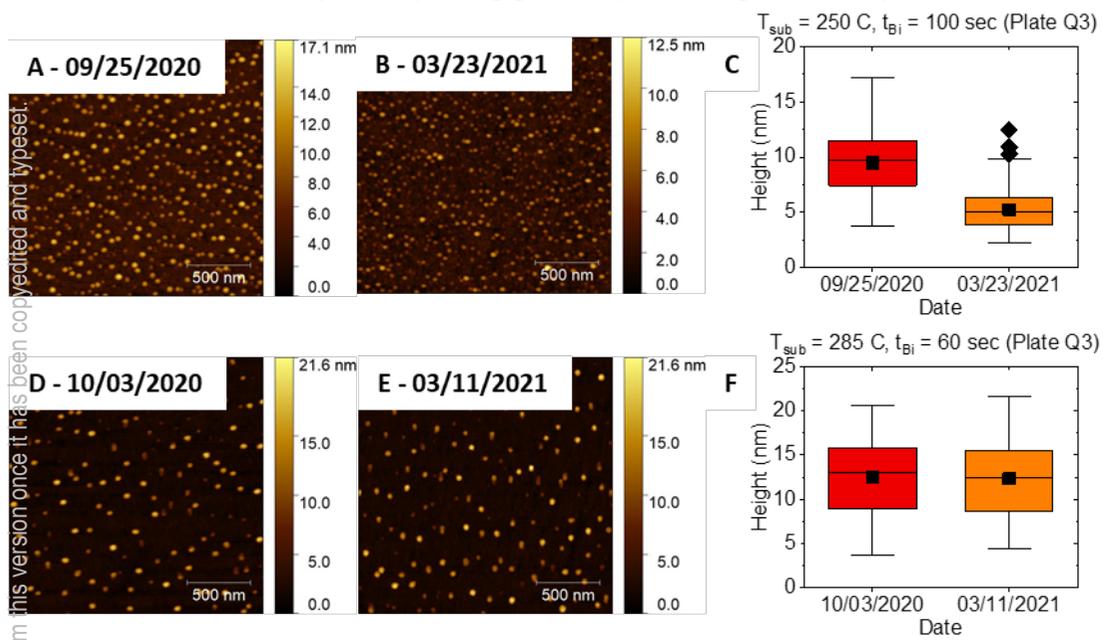

Attempts at repeating growths (different growth plate)

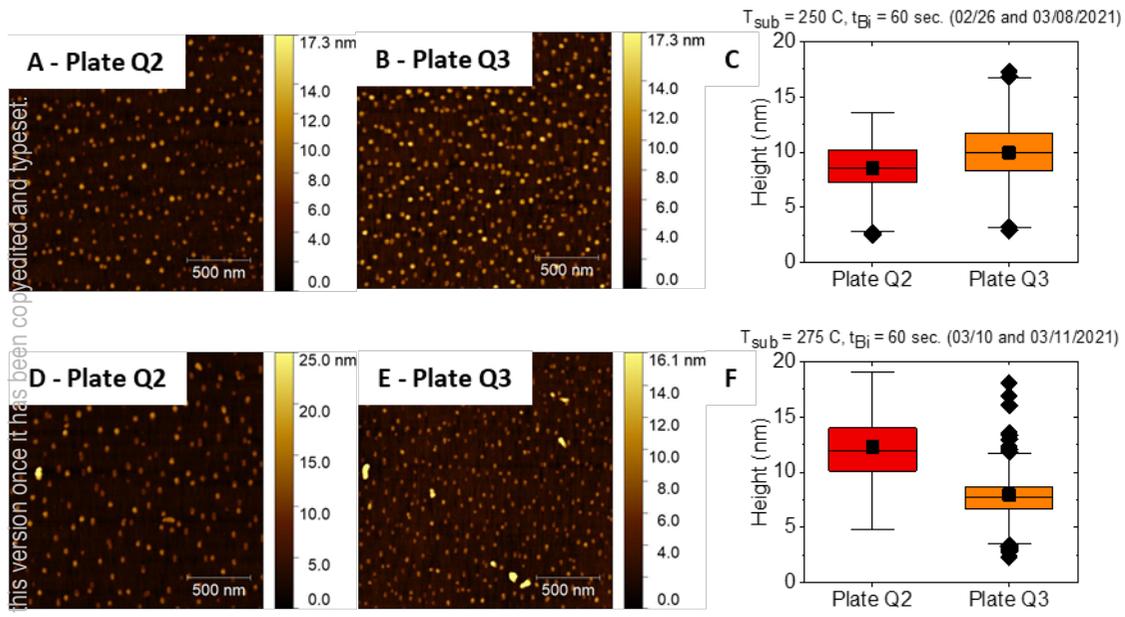

# Growth of nanoparticles on smoother GaAs

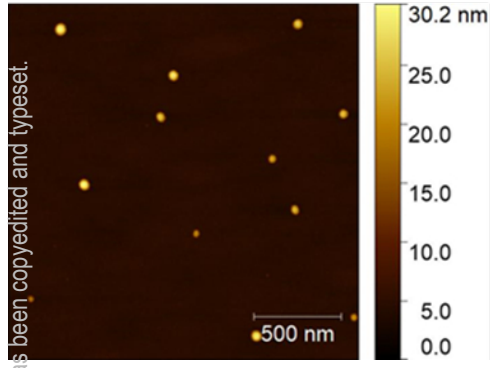
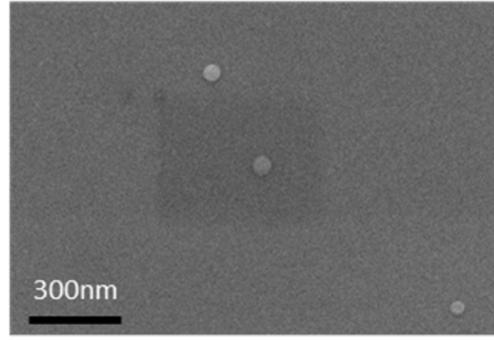